\documentclass[journal]{IEEEtran}
\usepackage[nobreak]{cite}
\usepackage{graphicx}
\usepackage{float}
\usepackage{amsmath}
\usepackage{amssymb}
\usepackage{amsfonts}
\usepackage[printonlyused]{acronym}
\usepackage[capitalise]{cleveref}
\crefformat{equation}{#2(#1)#3}
\Crefformat{equation}{#2(#1)#3}
\usepackage{tikz}
\usetikzlibrary{calc}
\usepackage{url}
\newcommand{\positiontextbox}[4][]{%
  \begin{tikzpicture}[remember picture,overlay]
    \node[inner sep=3pt, fill=yellow,align=left,draw,line width=1pt,#1] at ($(current page.north west) + (#2,-#3)$) {\parbox{.95\paperwidth}{#4}};
  \end{tikzpicture}%
}

\providecommand{\expe}[1]{\ensuremath{\mathrm{e}^{#1}}}
\def\sinc{\mathop{\mathrm{sinc}}\nolimits}

\begin{document}

\title{Time-Modulated Phased Array Controlled with Nonideal Bipolar Squared Periodic Sequences}	
\author{Roberto Maneiro-Catoira,~\IEEEmembership{Member,~IEEE,}
	Julio Br\'egains,~\IEEEmembership{Senior~Member,~IEEE,}\\
	Jos\'e A. Garc\'ia-Naya,~\IEEEmembership{Member,~IEEE,}
	and~Luis Castedo,~\IEEEmembership{Senior~Member,~IEEE}
		\thanks{%
		$^*$ Corresponding author: Jos\'e A. Garc\'ia-Naya (jagarcia@udc.es).
		
		This work has been funded by the Xunta de Galicia (ED431C 2016-045, ED341D R2016/012, ED431G/01), the Agencia Estatal de Investigaci\'on of Spain (TEC2015-69648-REDC, TEC2016-75067-C4-1-R) and ERDF funds of the EU (AEI/FEDER, UE).
		
		The authors are with the University of A Coru\~na, Spain. E-mail: roberto.maneiro@udc.es, julio.bregains@udc.es, jagarcia@udc.es, luis@udc.es
		
		Digital Object Identifier 10.1109/LAWP.2019.2892657}
	\thanks{}}

\markboth{IEEE ANTENNAS AND WIRELESS PROPAGATION LETTERS}%
{}
\maketitle

\acrodef{ADC}[ADC]{analog-to-digital converter}
\acrodef{AWGN}[AWGN]{additive white Gaussian noise}
\acrodef{ASK}[ASK]{amplitude-shift keying}
\acrodef{BER}[BER]{bit error ratio}
\acrodef{BF}[BF]{beamforming}
\acrodef{BS}[BS]{beam-steering}
\acrodef{BFN}[BFN]{beamforming network}
\acrodef{SER}[SER]{symbol error rate}
\acrodef{DAC}[DAC]{digital-to-analog converter}
\acrodef{ETMA}[ETMA]{enhanced time-modulated array}
\acrodef{DC}[DC]{direct current}
\acrodef{DOA}[DOA]{direction of arrival}
\acrodef{DSB}[DSB]{double sideband}
\acrodef{FSK}[FSK]{frequency-shift keying}
\acrodef{FT}[FT]{Fourier Transform}
\acrodef{ISI}[ISI]{inter-symbol interference}
\acrodef{HPBW}[HPBW]{half-power beam-width}
\acrodef{HT}[HT]{Hilbert Transform}
\acrodef{LMD}[LMD]{linearly modulated digital}
\acrodef{LNA}[LNA]{low noise amplifier}
\acrodef{MRC}[MRC]{maximum ratio combining}
\acrodef{MBPA}[MBPA]{Multibeam phased-array antenna}
\acrodef{MMIC}[MMIC]{monolithic microwave integrated circuit}
\acrodef{MSLL}[MSLL]{maximum side-lobe level}
\acrodef{NMLW}[NMLW]{normalized main-lobe width}
\acrodef{NPD}[NPD]{normalized power density}
\acrodef{PCB}[PCB]{printed circuit board}
\acrodef{PS}[PS]{phase shifter}
\acrodef{PSK}[PSK]{phase-shift keying}
\acrodef{QAM}[QAM]{quadrature amplitude modulation}
\acrodef{RF}[RF]{radio frequency}
\acrodef{RFC}[RFC]{Rayleigh fading channel}
\acrodef{RPDC}[RPDC]{reconfigurable power/divider combiner}
\acrodef{SA}[SA]{simulated annealing}
\acrodef{SPMT}[SPMT]{single-pole multiple-throw}
\acrodef{SPST}[SPST]{single-pole single-throw}
\acrodef{SLL}[SLL]{sideband-lobe level}
\acrodef{SR}[SR]{sideband radiation}
\acrodef{SNR}[SNR]{signal-to-noise ratio}
\acrodef{SPDT}[SPDT]{single-pole dual-throw}
\acrodef{SSB}[SSB]{single sideband}
\acrodef{SWC}[SWC]{sum of weighted cosines}
\acrodef{STA}[STA]{static array}
\acrodef{TM}[TM]{time-modulated}
\acrodef{TMA}[TMA]{time-modulated array}
\acrodef{TMPA}[TMPA]{time-modulated phased array}
\acrodef{VPS}[VPS]{variable phase shifter}
\acrodef{VGA}[VGA]{variable-gain amplifier}
\acrodef{UWB}[UWB]{ultra-wide band}
\begin{abstract}
Bipolar (+/-1) sequences with no zero state suit particularly well for safeguarding the switched feeding network efficiency when applied to time-modulated arrays (TMAs). During the zero state of a conventional time-modulating sequence, if a given array element is switched off, a certain amount of energy of the transmitted/received signal is wasted. We propose a novel single sideband time-modulated phased array (TMPA) architecture governed by realistic bipolar squared sequences in which the rise/fall time of the switches is considered. By using single-pole dual-throw switches and non-reconfigurable passive devices, the TMPA exploits, exclusively, the first positive harmonic pattern while exhibiting an excellent performance in terms of efficiency and control level of the undesired harmonics without using synthesis optimization algorithms (software simplicity).
\end{abstract}

\begin{IEEEkeywords}
Antenna arrays, time-modulated arrays, beamsteering.
\end{IEEEkeywords}

\acresetall

%
\IEEEpeerreviewmaketitle
	
\positiontextbox{11cm}{27cm}{\footnotesize \textcopyright 2019 IEEE. This version of the article has been accepted for publication, after peer review. Personal use of this material is permitted. Permission from IEEE must be obtained for all other uses, in any current or future media, including reprinting/republishing this material for advertising or promotional purposes, creating new collective works, for resale or redistribution to servers or lists, or reuse of any copyrighted component of this work in other works. Published version:
\url{10.1109/LAWP.2019.2892657}}

\section{Introduction}
\IEEEPARstart{P}{hased} array architectures are based, in general, on \acp{VPS}. Unfortunately, such devices still exhibit a number of inconveniences such as cost, insertion losses, or phase resolution \cite{Qorvo,ManeiroCatoira_2018_ACCESS}. The application of switched \acp{TMA} to \ac{BS} is attractive in terms of cost and simplicity. However, their efficiency and flexibility must be improved and, particularly, the following concerns are identified in the literature:
1) The lack of efficiency in the distribution of the spectral energy among the working harmonics of rectangular pulses for multiple \ac{BS} \cite{Maneiro2017_a}.
2) The overlooking of the presence of mirror-frequency diagrams at the negative harmonics \cite{Maneiro2017_a, Bogdan2016,Poli2011}.
3) The proportionality between the phases of harmonics limits the multiple \ac{BS} flexibility \cite{Maneiro2017_b}.
4)  The fundamental mode beam has no scanning ability \cite{Maneiro2017_a,Poli2011}, unless supplementary delay lines are included \cite{Sun2016}.
5) Defining the overall time-modulation efficiency as $\eta =\eta_\text{TMA}\cdot \eta_s$, then $\eta_{\text{TMA}}=P_U^{\text{TM}}/P_R^{\text{TM}}$ ($P_U^{\text{TM}}$ and $P_R^{\text{TM}}$ are the useful and total mean power values radiated by the \ac{TMA}, respectively) accounts for the ability of the \ac{TMA} to filter out and radiate only the useful harmonics and $\eta_s=P_R^{\text{TM}}/P_R^{\text{ST}}$ ($P_R^{\text{ST}}$ is the total mean power radiated by a uniform static array with $N$ elements) accounts for the reduction of the total mean power radiated by a uniform static array caused by the insertion of the \ac{TMA} switched feeding network. In this respect, the most common is either the switched feeding network efficiency $\eta_s$ is contemplated but without providing the \ac{TMA} efficiency $\eta_\text{TMA}$ \cite{Bogdan2016}, or vice versa \cite{Amin_Yao2015}. Hence, the efficiency analyses available in the literature are either incomplete or misleading.

Given these shortcomings, in this work we focus our efforts on the design of a switched \ac{TMPA} performing \ac{BS} over a single harmonic.

The main contributions of this work are:
1) The modeling of a \ac{SSB} switched \ac{TMPA} architecture governed by bipolar sequences generated by means of \ac{SPDT} switches to efficiently perform \ac{BS}. 
2) The analysis of the impact of the rise/fall time of the \ac{SPDT} switches on both efficiency factors ($\eta_{\text{TMA}}$ and $\eta_s$) and on the control level of the undesired harmonics.

\section{Design of the \ac{SSB} \ac{TMPA} }\label{sec:mathematicalAnalysis}

Any arbitrary periodic waveform (e.g., a square wave) can be expressed, through the trigonometric Fourier series expansion, as an infinite sum (or linear combination) of simple sine and cosine waves. In this letter, however, our aim is to approximate ---with a finite sum, see \cref{eq:w(t) sin} in the forthcoming analysis--- a sine wave using a linear combination of non-sinusoidal bipolar square waves\footnote{We consider non-ideal bipolar waves, which are properly trapezoidal, to model a realistic behavior of the physical switches employed to generate such square waves. The impact of pulse shaping on the sideband radiation of switched \acp{TMA} was addressed for the first time in \cite{Bekele2013} and next in \cite{Amin_Yao2015}.}. In this way, the global idea of this letter is to employ simple time-delayed sine waves (more precisely, good approximations of sine waves) to modulate each antenna array excitation to perform \ac{BS}. Notice that each sine wave will be synthesized as a linear combination of bipolar square waves which are easily generated by means of \ac{SPDT} switches.

\begin{figure}[t]
	\centering
	\includegraphics[width=0.9\columnwidth]{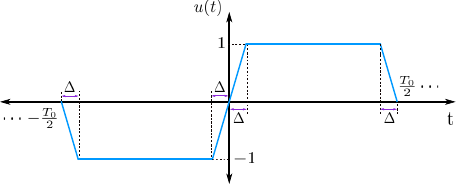}
	\caption{Periodic bipolar odd pulse. $\Delta$ is the rise/fall time of the switches.} 
	\label{fig:basic bipolar pulse}
\end{figure}

First, let us study (see \cref{fig:basic bipolar pulse}) the basic periodic ($T_0$) bipolar pulse $u(t)$ considered in this work. This pulse models a realistic bipolar $\pm$1 square periodic pulse, where $\Delta$ is the rise/fall time of the switches used to implement such a pulse (hence, from a pure mathematical point of view, they are properly trapezoidal if $\Delta \neq 0$). Its trigonometric Fourier coefficients are:
\begin{align}\label{eq:Fourier coefficients u(t)}
U_{q}=\frac{4}{T_0}\int_{0}^{T_0/2}u(t)\sin(q\omega_0t)dt=\begin{cases}
\frac{4\sinc(q\omega_0 \Delta)}{\pi q} & q \hspace{0.2cm}\text{odd}\\
0&q \hspace{0.2cm}\text{even},
\end{cases}
\end{align}
with $\omega_0=2\pi/T_0$ and, thus, we can write $u(t)$ as
\begin{equation}\label{eq:u(t) sin}
u(t)=\frac{4}{\pi}\sum_{q = 1,3,5, \dots}^\infty\frac{\sinc(q\omega_0 \Delta)}{q}\sin(q\omega_0t).
\end{equation}
If we consider a signal $v(t)$ with exactly the same characteristics of $u(t)$ but with triple fundamental frequency, we can express
\begin{equation}\label{eq:v(t) sin}
v(t)=\frac{4}{\pi}\sum_{q = 1,3,5, \dots}^\infty\frac{\sinc(q(3\omega_0) \Delta)}{q}\sin(q(3\omega_0)t),
\end{equation}
and we can approximate our simple sine function with fundamental frequency $\omega_0$ by means of the linear combination of bipolar periodic signals, i.e.,
\begin{equation}\label{eq:w(t) sin}
w(t)=u(t)-1/3v(t)=\frac{4}{\pi}\sum_{q \in \Upsilon}\frac{\sinc(q\omega_0 \Delta)}{q}\sin(q\omega_0t),
\end{equation}
with $\Upsilon=\{q\in \mathbb{N^*}/ q\text{ odd}; q\neq   \mathring{3}\}=\{1, 5, 7, 11, \dots\}$, hence removing the frequencies that are multiple of 3. Additionally, to steer the exploited harmonic pattern of the \ac{TMA}, we must consider a time-shifted version of $w(t)$, $w_n(t)= w(t-D_n)$, with $D_n$ being the corresponding time-delay variable,  hence
\begin{equation}\label{eq:w_n(t) sin}
w_n(t)=\frac{4}{\pi}\sum_{q \in \Upsilon}\frac{\sinc(q\omega_0 \Delta)}{q}\sin(q\omega_0(t-D_n)).
\end{equation}

\begin{figure}[t]
\centering
\includegraphics[width=0.9\columnwidth]{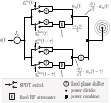}
\caption{Scheme of the feeding for the $n$-th antenna element of the proposed
\ac{SSB} \ac{TMPA}. Notice that $g_n^{\omega_0}(t)$ and $g_n^{3\omega_0}(t)$ are the corresponding unipolar versions of $u_n(t)$ and $v_n(t)$, which are bipolar. } 
\label{fig:n-th element TMPA}
\end{figure}

Let us now consider a linear \ac{TMPA} with $N$ isotropic elements with unitary static excitations $I_n = 1$, $n \in \{0, 1,\dots, N-1\}$ whose $n$-th element feeding scheme is shown in \cref{fig:n-th element TMPA}. Such a feeding network has a two-branch structure according to the time-modulating waveform ($w_n(t)$ or $w_n(t-\tau)$) ---with $\tau$ being a previously fixed time delay--- followed by a $\pi/2$ fixed phase shifting. 
According to \cref{fig:n-th element TMPA}, the time-varying array factor is given by
\begin{equation}\label{eq:array factor time-domain}
F(\theta,t)=\sum_{n=0}^{N-1}\left[\frac{w_n(t)}{\sqrt{2}}+j\frac{w_n(t-\tau)}{\sqrt{2}}\right]\expe{jkz_n\cos\theta},
\end{equation} 
where $z_n$ represents the $n$-th array element position on the $z$ axis, $\theta$ is the angle with respect to such a main axis, and $k=2\pi/\lambda$ represents the wavenumber for a carrier wavelength $\lambda = 2\pi \mathrm{c} /\omega_c$. We begin the analysis of \cref{eq:array factor time-domain} by evaluating the term $w_n(t)+jw_n(t-\tau)$. For the sake of simplicity, we will analyze the \ac{FT} of such a term, i.e., $\mathrm{FT}[w_n(t)]+j\mathrm{FT}[w_n(t-\tau)]$ where, by virtue of \cref{eq:w_n(t) sin}, 
\begin{align}\label{eq:pulses_frequency_domain}
& \mathrm{FT}[w_n(t)] = \frac{4}{j}\sum_{q\in\Upsilon}\frac{\sinc(q\omega_0 \Delta)}{q}\left[\expe{-jq\omega_0D_n}\delta(\omega-q\omega_0)-\right.\notag\\
&\left. -\expe{jq\omega_0D_n}\delta(\omega+q\omega_0)\right], \text{ and}\notag\\
&\mathrm{FT}[w_n(t-\tau)] = \expe{-j\omega \tau}\mathrm{FT}[w_n(t)] =\notag\\ &=\frac{4}{j}\sum_{q\in\Upsilon}\frac{\sinc(q\omega_0 \Delta)}{q}\left[\expe{-jq\omega_0 \tau}\expe{-jq\omega_0D_n}\delta(\omega-q\omega_0)-\right.\notag\\
&\left.-\expe{jq\omega_0 \tau}\expe{jq\omega_0 D_n}\delta(\omega+q\omega_0)\right].
\end{align} 
If we select a delay $\tau$ verifying that $\omega_0\tau=\pi/2$, then $\expe{-jq\omega_0\tau}=(-j)^q$ and $\expe{jq\omega_0\tau}=j^q$, and hence we have that
\begin{align}\label{eq:suma de wn y delayed frequency_domain}
&\mathrm{FT}[w_n(t)]+j\mathrm{FT}[w_n(t-\tau)]=\notag\\
&=\frac{4}{j}\sum_{q\in\Upsilon}\frac{\sinc(q\omega_0 \Delta)}{q}\Big[(1-(-j)^{q+1})\expe{-jq\omega_0D_n}\delta(\omega-q\omega_0)+\notag\\
&+(-1-j^{q+1})\expe{jq\omega_0D_n}\delta(\omega+q\omega_0)\Big].
\end{align}
Considering the sets of indexes $\Upsilon_1=\{q=4k-3; k\in \mathbb{N^*}; q\neq   \mathring{3}\}=\{1, 5, 13, 17, \dots\}$ and $\Upsilon_2=\{q=4k-1; k\in \mathbb{N^*}; q\neq   \mathring{3}\}=\{7, 11, 19, 23 \dots\}$, such that $\Upsilon=\Upsilon_1\cup\Upsilon_2$, then
\begin{align}\label{eq:valores  (1 + (-j)^{q+1})...}
1-(-j)^{q+1}=\begin{cases}
2 & q\in \Upsilon_1\\
0&q\in \Upsilon_2,
\end{cases};\  
-1-j^{q+1}=\begin{cases}
-2 & q\in \Upsilon_2\\
0&q\in \Upsilon_1.
\end{cases}
\end{align}
Hence, we can rewrite \cref{eq:suma de wn y delayed frequency_domain} as
\begin{align}\label{eq:suma de wn y delayed frequency_domain 2}
&\mathrm{FT}[w_n(t)]+j\mathrm{FT}[w_n(t-\tau)]=\notag\\
&=\frac{8}{j}\sum_{q\in\Upsilon_1}\frac{\sinc(q\omega_0 \Delta)}{q}\expe{-jq\omega_0D_n}\delta(\omega-q\omega_0)+\notag\\
&+\frac{(-8)}{j}\sum_{q\in\Upsilon_2}\frac{\sinc(q\omega_0 \Delta)}{q}\expe{jq\omega_0D_n}\delta(\omega+q\omega_0),
\end{align}
and we realize that the harmonics with orders $-1,-5,7, 11, \dots$ are removed. By applying the inverse \ac{FT} to \cref{eq:suma de wn y delayed frequency_domain 2}, we have
\begin{align}\label{eq:suma de wn y delayed time domain plus}
&w_n(t)+jw_n(t-\tau)=\sum_{q\in\Upsilon_1}\frac{4\sinc(q\omega_0 \Delta)}{j\pi q}\expe{-jq\omega_0D_n}\expe{jq\omega_0t}+\notag\\
&+\sum_{q\in\Upsilon_2}\frac{-4\sinc(q\omega_0 \Delta)}{j\pi q}\expe{jq\omega_0D_n}\expe{-jq\omega_0t},
\end{align}
and \cref{eq:array factor time-domain} can be rewritten as
\begin{align}\label{eq:F time plus}
&F(\theta,t)=\frac{1}{\sqrt{2}}\sum_{n=0}^{N-1}\left[\sum_{q\in\Upsilon_1}\frac{4\sinc(q\omega_0 \Delta)}{j\pi q}\expe{-jq\omega_0D_n}\expe{jq\omega_0t}+\right.\notag\\
&\left.+\sum_{q\in\Upsilon_2}\frac{-4\sinc(q\omega_0 \Delta)}{j\pi q}\expe{jq\omega_0D_n}\expe{-jq\omega_0t}\right]\expe{jkz_n\cos\theta}.
\end{align} 
\begin{figure}[t]
	\centering
	\includegraphics[width=0.9\columnwidth]{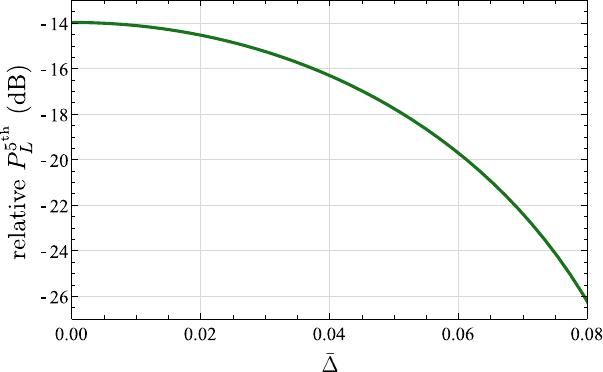}
	\caption{Representation of  the difference between the peak levels of the $1^{\text{st}}$ and $5^{\text{th}}$ harmonics diagrams ($P_L^{5^\text{th}}$) as a function of $\bar{\Delta}$ \cref{eq:PL5}. The plot reveals that the peak of the most meaningful undesired harmonic in the proposed \ac{TMPA} is very sensitive to $\bar{\Delta}$.} 
	\label{fig:relative peak level}
\end{figure}
We now define
\begin{align}\label{eq:definition of F1 y F2}
&F_1(\theta,t)_q=\expe{jq\omega_0t}\sum_{n=0}^{N-1} I_{nq}\expe{jkz_n\cos\theta},\text{ and}\notag\\
&F_2(\theta,t)_q=\expe{-jq\omega_0t}\sum_{n=0}^{N-1} I_{nq}^{'}\expe{jkz_n\cos\theta},
\end{align} 
to finally obtain
\begin{equation}\label{eq:array factor3}
F(\theta,t)=\sum_{q\in\Upsilon_1}F_1(\theta,t)_q+\sum_{q\in\Upsilon_2}F_2(\theta,t)_q, 
\end{equation}	
where the dynamic excitations  $I_{nq}$ and $I_{nq}^{'}$ are given by
\begin{align}\label{eq:dynamic excitations}
&I_{nq}=\frac{4\sinc(q\omega_0 \Delta)}{j\pi \sqrt{2}q}\expe{-jq\omega_0D_n}, \hspace{0.2cm}q\in \Upsilon_1,\text{ and}\notag\\
&I_{nq}^{'}=\frac{-4\sinc(q\omega_0 \Delta)}{j\pi \sqrt{2}q}\expe{jq\omega_0D_n}, \hspace{0.2cm}q\in \Upsilon_2,
\end{align}

\section{Impact of $\Delta$ on the \ac{TMPA} performance}\label{sec:efficiency}
In this section, we analyze 1) the effects produced by $\Delta$ on the peak level of the power radiated pattern at the most meaningful undesired harmonic (i.e., $|F_1(\theta,t)_5|^2$) with respect to the peak level at the exploited harmonic ($|F_1(\theta,t)_1|^2$); and 2) the impact of $\Delta$ on the efficiency of the time modulation operation applied to the \ac{TMPA}.
Regarding the first aim, according to \cref{eq:dynamic excitations}, and by considering the normalized rise/fall time of the switches $\bar{\Delta}=\Delta/T_0$, we define the figure of merit 
\begin{equation}\label{eq:PL5}
P_L^{5^{\text{th}}}(\text{dB})=20\log_{10}\left|\frac{I_{n5}}{I_{n1}}\right|=20\log_{10}\left|\frac{\sinc(10\pi \bar{\Delta})}{5\sinc(2\pi \bar{\Delta})}\right|,
\end{equation}
which quantifies the difference, in dB, between the peak levels of the $1^{\text{st}}$ and $5^{\text{th}}$ harmonics diagrams. The representation of $P_L^{5^\text{th}}$ as a function of $\bar{\Delta}$ ---according to \cref{eq:PL5}--- is shown in \cref{fig:relative peak level}, which illustrates that it is possible to keep $P_L^{5^\text{th}}$ below a predetermined threshold by selecting an appropriate value of $\bar{\Delta}$. Hence, $\bar{\Delta}$ manifests itself as a crucial design parameter.

\begin{figure}[t]
	\centering
	\includegraphics[width=0.9\columnwidth]{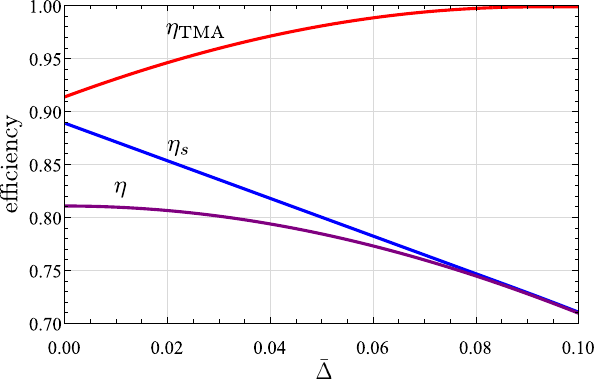}
	\caption{Efficiencies of the \ac{TMPA} as a function of the normalized rise/fall time of the \ac{SPDT} switches $\bar{\Delta}$.} 
	\label{fig:efficiencies}
\end{figure}

However, $\bar{\Delta}$ also impacts on the efficiency of the time modulation operation in the \ac{TMPA}. As described in the introduction, such an efficiency depends on two factors: $\eta_{\text{TMA}}$ and $\eta_s$. Hence, the selection of $\bar{\Delta}$ impacts not only on $\eta_{\text{TMA}}$ (as described in \cite{Amin_Yao2015} using other kind of pulses and hardware), but also (and this constitutes a key contribution of this work) on $\eta_s$ and, consequently, on the overall efficiency $\eta$. 

\begin{figure*}[!t]
	\centering
	\includegraphics[width=14cm]{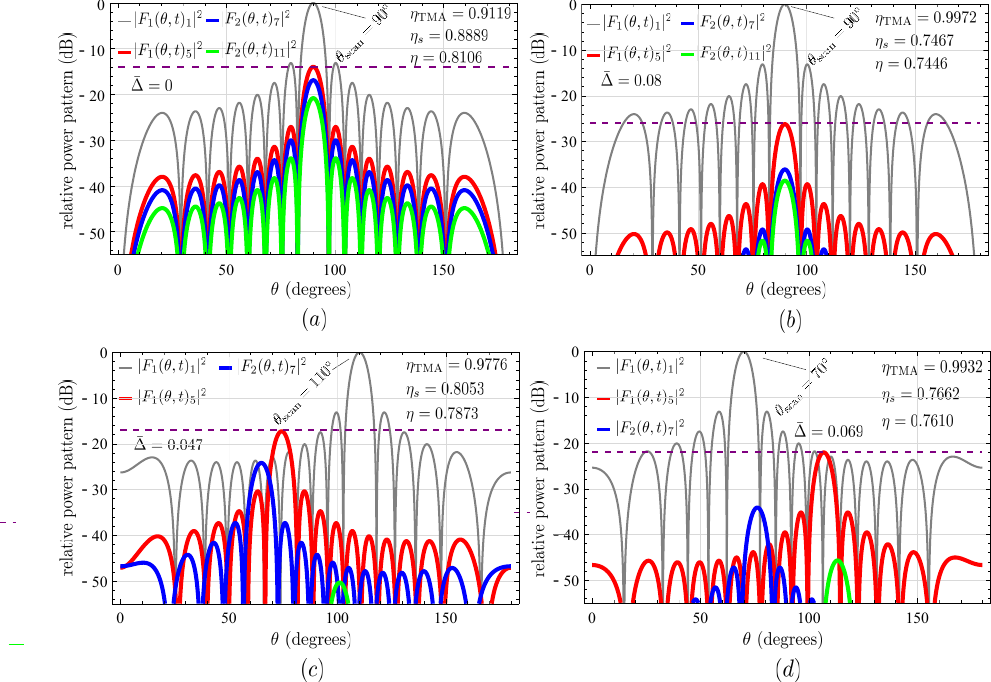}
	\caption{Relative power radiated patterns of the proposed \ac{TMPA} with $N=16$ for different normalized rise/fall time of the \ac{SPDT} switches $\bar{\Delta}$ and time delays $D_n$. The scenarios (a) to (d) are explained in \cref{sec:numerical}.} 
	\label{fig:radiated patterns}
\end{figure*}

We next accurately characterize the variation of both terms of $\eta$ with $\Delta$. To simplify the analysis, but without any relevant loss of generality, we consider a uniform linear array with inter-element distance of $\lambda/2$ transmitting a single carrier with normalized power. We start by quantifying $P_R^{\text{TM}}$, the total mean radiated power by the \ac{TMPA}, which is given by the expression $P_R^{\text{TM}}=\sum_{q\in\Upsilon_1}(p_1)_q+\sum_{q\in\Upsilon_2}(p_2)_q$, see \cite[(16)]{Maneiro2017_a}, being $(p_1)_q$ and $(p_2)_q$ the total mean radiated power values at the frequencies $\omega_c+q\omega_0$ with $q \in\Upsilon_1$ and $\omega_c-q\omega_0$ with $q \in\Upsilon_2$, respectively. As $(p_1)_q=4\pi\sum_{n=0}^{N-1}|I_{nq}|^2$ and $(p_2)_q=4\pi\sum_{n=0}^{N-1}|I_{nq}^{'}|^2$, see \cite{Maneiro2017_a}, and 
$|I_{nq}|^2=|I_{nq}^{'}|^2=8\sinc^2(q\omega_0\Delta)/(\pi^2q^2)$ from \cref{eq:dynamic excitations}, we arrive at $P_R^{\text{TM}}=\frac{32N}{\pi}\sum_{q\in\Upsilon}\frac{\sinc^2(q\omega_0\Delta)}{q^2}$. Notice that the useful mean radiated power is  $P_U^{\text{TM}}=(p_1)_1=4\pi\sum_{n=0}^{N-1}|I_{n1}|^2=32N\sinc^2(\omega_0\Delta)/\pi$. On the other hand, the total mean power radiated by a uniform static array with $N$ elements, $P_R^{ST}$, is calculated as the total mean transmitted power over the array factor $F^{\text{ST}}(\theta)=\sum_{n=0}^{N-1}\expe{jkz_n\cos\theta}$ and, hence, $P_{\text{R}}^{\text{ST}}= \int_{0}^{2\pi}\int_{0}^{\pi} |F^{\text{ST}}(\theta)|^2\sin(\theta)d\theta d\varphi=4\pi N$. Therefore:
\begin{align}
\eta_{\text{TMA}}(\bar{\Delta}) &= \frac{P_U^{\text{TM}}}{P_R^{\text{TM}}}=\frac{\sinc^2(2\pi\bar{\Delta})}{\sum_{q\in\Upsilon}\frac{\sinc^2(2\pi q\bar{\Delta})}{q^2}}, \text{ and} \label{eq:eta TMA}\\
\eta_s(\bar{\Delta}) &= \frac{P_R^{\text{TM}}}{P_R^{\text{ST}}}=\frac{8}{\pi^2}\sum_{q\in\Upsilon}\frac{\sinc^2(2\pi q\bar{\Delta})}{q^2}\label{eq:eta s},
\end{align}
and consequently,
\begin{equation}\label{eq:eta}
\eta(\bar{\Delta})=\eta_{\text{TMA}}(\bar{\Delta}) \cdot \eta_s(\bar{\Delta})= \frac{8}{\pi^2}\sinc^2(2\pi \bar{\Delta}).
\end{equation}

\cref{fig:efficiencies} illustrates the different \ac{TMPA} efficiencies as a function of $\bar{\Delta}$. From \cref{fig:relative peak level,fig:efficiencies} we observe that the better the $P_L^{5^\text{th}}$, the better the $\eta_{\text{TMA}}$, as expected. We also appreciate the opposite slopes of both factors of the efficiency. Finally, we highlight the existence of a trade-off between $P_L^{5^\text{th}}$ and $\eta$ when  $\bar{\Delta}$ changes. Such a factor should be accurately taken into account by the antenna designer. 

\section{Numerical simulations}\label{sec:numerical}
In this section, we show the behavior of the proposed \ac{TMPA} by means of several numerical examples. In line with \cref{fig:relative peak level,fig:efficiencies}, \cref{fig:radiated patterns}a shows that when $\bar{\Delta}=0$ (ideal squared pulses), $P_L^{5^\text{th}}$ ($-14$\,dB) and $\eta_{\text{TMA}}$ ($91$\%) exhibit their worst values while $\eta_s$ ($89$\%) and $\eta$ ($81$\%) achieve their best values, and the directivity  $G_D = 11.64$\,dBi. Notice that the time-delays $D_n$ are set to zero, and hence the scanning angle of all patterns (see \cref{eq:F time plus}) will be $\theta_{\text{scan}}=90^\circ$.

In \cref{fig:radiated patterns}b, without varying $\theta_{\text{scan}}$, $\bar{\Delta}$ is raised to $0.08$ and we observe that $G_D=12.03$\,dBi and both $P_L^{5^\text{th}}$ ($-26$\,dB) and $\eta_{\text{TMA}}$ are sensibly improved ($87.7$\% and $9.35$\%, respectively) at the expense of a worse $\eta_s$ (decreased to $16.0$\%), and leading also to a worse $\eta$ (a reduction of $8.14$\%).

Figs.~\ref{fig:radiated patterns}c and \ref{fig:radiated patterns}d show both the scanning ability of the \ac{TMPA} and the trade off between $\eta$ and   $P_L^{5^\text{th}}$ when different values of $\bar{\Delta}$ are selected. More specifically, the $P_L^{5^\text{th}}$  threshold  in \cref{fig:radiated patterns}c is fixed at $-17$\,dB and, according to \cref{eq:PL5} and \cref{fig:relative peak level}, $\bar{\Delta}=0.047$, thus achieving an $\eta$ that is only $2.9$\% below its maximum value while $G_D=11.94$\,dBi. In this case, $D_n$ are selected to accomplish a $\theta_{\text{scan}}=110^\circ$ by simply assigning progressive phases to the array elements, i.e., considering $D_{n}/T_0=n \cos(\theta_\text{scan})$. In \cref{fig:radiated patterns}d, the $P_L^{5^\text{th}}$ threshold is fixed at a more stringent level, $-22$\,dB, which leads to $\bar{\Delta}=0.069$, thus achieving an $\eta$ which is $6.1$\% below its maximum value. In this case, $D_n$ are selected to accomplish a $\theta_{\text{scan}}=70^\circ$ while $G_D = 12.01$\,dBi. Notice that, since all the excitations of the patterns are uniform, the \ac{SLL} and the \ac{HPBW} only depend on $N$.

\section{Conclusions}

We proposed a novel \ac{TMPA} scheme based on time modulation with  non-ideal bipolar squared periodic sequences. When the rise/fall time of the pulses changes, we have accurately analyzed the trade off between the two components of the time modulation efficiency, as well as the trade off between such an efficiency and the peak level of the most meaningful undesired harmonic. Accordingly, the structure presents the advantage that the designer can select an adequate  rise/fall time of the bipolar pulses (and therefore, a particular \ac{SPDT} switch) in order to guarantee a certain threshold for the undesired harmonics while the efficiency of the time modulation remains above the value dictated by the required performance level.
  
\vfill





\bibliographystyle{IEEEtran}
%
\bibliography{IEEEabrv,main}

%



\end{document}